\documentclass[ reprint,
 amsmath,amssymb,
 aps,
floatfix,
]{revtex4-2}
\usepackage{graphicx}% Include figure files
\usepackage{dcolumn}% Align table columns on decimal point
\usepackage{bm}% bold math

\usepackage{caption}
\usepackage{xcolor}
\usepackage{hyperref}

\begin{document}

\preprint{}

\title{Insights into Polycrystalline Microstructure of Blood Films with 3D Mueller Matrix Imaging Approach}% Force line breaks with \\

\author{Volodimyr A. Ushenko}
 \affiliation{Optics and Publishing Department, Chernivtsi National University, 2 Kotsiubynskyi Str., Chernivtsi, 58012, Ukraine}%Lines break automatically or can be forced with \\

\author{Anton Sdobnov}
 \affiliation{Optoelectronics and Measurement Techniques, University of Oulu, P.O. Box 4500, Oulu, FI-90014, Finland}%Lines break automatically or can be forced with \\

\author{Liliya Trifonyuk}
 \affiliation{Rivne State Medical Center, 78 Kyivska Str., Rivne, 33007, Ukaraine}%Lines break automatically or can be forced with \\

 \author{Alexander V. Dubolazov}
 \affiliation{Optics and Publishing Department, Chernivtsi National University, 2 Kotsiubynskyi Str., Chernivtsi, 58012, Ukraine}%Lines break automatically or can be forced with \\

  \author{Alexander Doronin}
 \affiliation{School of Engineering and Computer Science, Victoria University ofWellington, 6140 Wellington, New Zealand}%Lines break automatically or can be forced with \\

\author{Yuriy A. Ushenko}
 \affiliation{Optics and Publishing Department, Chernivtsi National University, 2 Kotsiubynskyi Str., Chernivtsi, 58012, Ukraine}%Lines break automatically or can be forced with \\

 \author{Irina V. Soltys}
 \affiliation{Optics and Publishing Department, Chernivtsi National University, 2 Kotsiubynskyi Str., Chernivtsi, 58012, Ukraine}%Lines break automatically or can be forced with \\

  \author{Mykhailo P. Gorsky}
 \affiliation{Optics and Publishing Department, Chernivtsi National University, 2 Kotsiubynskyi Str., Chernivtsi, 58012, Ukraine}%Lines break automatically or can be forced with \\

  \author{Alexander G. Ushenko}
 \affiliation{Optics and Publishing Department, Chernivtsi National University, 2 Kotsiubynskyi Str., Chernivtsi, 58012, Ukraine}%Lines break automatically or can be forced with \\
  \affiliation{College of Electrical Engineering, Taizhou Research Institute, Zhejiang University, Hangzhou 310027, China}

\author{Vyacheslav K. Gantyuk}
 \affiliation{Optics and Publishing Department, Chernivtsi National University, 2 Kotsiubynskyi Str., Chernivtsi, 58012, Ukraine}%Lines break automatically or can be forced with \\

\author{Wenjun Yan}
 \affiliation{College of Electrical Engineering, Taizhou Research Institute, Zhejiang University, Hangzhou 310027, China}

 \author{Alexander Bykov}
 \affiliation{Optoelectronics and Measurement Techniques, University of Oulu, P.O. Box 4500, Oulu, FI-90014, Finland}%Lines break automatically or can be forced with \\

\author{Igor Meglinski}
 \email{Correspondence: i.meglinski@aston.ac.uk}
 \affiliation{College of Engineering and Physical Sciences, Aston University, Birmingham, B4 7ET, UK}%Lines break automatically or can be forced with \\

\date{\today}% It is always \today, today,

\begin{abstract}

We introduce a 3D Mueller Matrix (MM) image reconstruction technique using digital holographic approach for the  layer-by-layer profiling thin films with polycrystalline structures, like dehydrated blood smears. The proposed method effectively extracts optical anisotropy parameters for a detailed quantitative analysis. The investigation revealed the method’s sensitivity to subtle changes in optical anisotropy properties resulting from alterations in the quaternary and tertiary structures of blood proteins, leading to disturbances in crystallization structures at the macro level at the very early stage of a disease. Spatial distributions of linear and circular birefringence and dichroism are analyzed in partially depolarizing polycrystalline blood films obtained from healthy tissues and cancerous prostate tissues at various stages of adenocarcinoma. Changes in the values of the 1st to 4th order statistical moments, characterizing the distributions of optical anisotropy in different ``phase" sections of the smear volumes, are observed and quantified. Comparative analysis of optical anisotropy distributions from healthy patients highlighted the 3rd and 4th order statistical moments for linear and circular birefringence and dichroism as the most promising for diagnostic purposes. We achieved an excellent accuracy ($>90\%$) for early cancer diagnosis and differentiation of its stages, demonstrating the technique's significant potential for rapid and accurate definitive cancer diagnosis compared to existing screening approaches.
\end{abstract}

\maketitle

%\tableofcontents

\section{\label{sec:Introduction}Introduction}

Over the past decades, there have been extensive studies on the formation of complex patterns arising during the evaporation of liquid droplets~\cite{PhysRep}.  Distinct patterns, including coffee rings~\cite{Deegan}, cracking~\cite{Iqbal}, and gelation~\cite{CHEN2016}, have been observed in biofluid droplets during drying. These patterns hold potential as straightforward diagnostic tools for assessing the health of both humans and livestock~\cite{Hertaeg,Cameron}. 

The dried blood droplet displays a discernible structure comprising three zones with varying thickness~\cite{Killeen}. Peripheral zone, characterized by a polycrystalline layer of albumin exhibiting linear birefringence and dichroism. Transitional zone, consisting of external and internal layers of optically isotropic cubic crystals of Na-Cl salt, with an intermediate layer of globulin demonstrating circular birefringence and dichroism. Central zone, featuring an external layer of cubic crystals of Na-Cl salt.
All zones contain multiple-scattering optical radiation elements—erythrocytes, platelets, and leukocytes—with manifestations of circular birefringence and dichroism~\cite{Brutin}. Alterations in the cellular and macromolecular constituents of blood, induced by diseases, are believed to contribute to variations in the dried drop patterns of both plasma and whole blood~\cite{Hertaeg,Cameron}. 

Spectroscopic techniques, including Raman, surface-enhanced Raman spectroscopy (SERS), infrared (IR), Fourier Transform IR (FTIR), and vibrational spectroscopy, have demonstrated the ability to characterize biomolecular presence and generate a biochemical fingerprint, offering implications for indicating disease states through the detection of protein imbalances within the drop during liquid evaporation~\cite{Cameron}. 
Alternatively to the spectroscopic techniques, the study explores spatially non-uniform, optically anisotropic biological structures with multiple-scattering layers using the Mueller matrix (MM)-based polarimetry approach~\cite{ushenko20183d,ushenko20213d,ushenko2021embossed,sdobnov2023polarization}. This method extracts mediated information, represented by 16 matrix elements, and integrates it to provide a comprehensive understanding of the polycrystalline structure within the biological layer, covering all scattering (depolarizing) inhomogeneities throughout the volume.

In contradistinction to conventional tissue specimens, anisotropic structures, exemplified by dry blood smears, offer a readily accessible alternative that eliminates the necessity for invasive biopsy procedures. The optical scrutiny of blood smears emerges as a promising and expeditious screening modality, particularly in the context of conditions such as prostate cancer, which instigates discernible alterations in the optical anisotropy characteristics~\cite{ushenko20183d,peyvasteh20203d}. This MM polarimetry approach presents a non-traumatic and straightforward methodology for screening purposes.

For practical clinical applications, it is crucial to extend MM polarimetry diagnostic techniques to assess the 3D polycrystalline structures of biological layers characterized by varying light scattering multiplicities and distinctive depolarizing capabilities. The layer-by-layer mapping of individual elements within the MM, specifically characterizing the parameters of phase and amplitude anisotropy, holds the potential to furnish critical diagnostic information. Achieving this goal involves the integration of previously reported differential MM techniques~\cite{ortega2011depolarizing,ortega2011mueller,ossikovski2014statistical,ossikovski2014general,devlaminck2013physical,devlaminck2014uniqueness} with holographic mapping of phase-inhomogeneous layers~\cite{ushenko20183d,peyvasteh20203d}.

The layer-by-layer distributions of depolarization degree, when aggregated, offer a thorough three-dimensional representation of depolarization and individual anisotropy parameters at a local scale. Preliminary investigations utilizing this methodology have unveiled a correlation between tissue features and three-dimensional Mueller matrix (3D MM) imaging, as evidenced in prior research~\cite{ushenko2021embossed}. This correlation establishes the groundwork for an effective and highly accurate differential diagnosis of prostate tumor tissues~\cite{ushenko20213d}.

In current study, we introduce the 3D MM mapping technique employing digital holographic reconstruction for the layer-by-layer profiling of partially depolarizing dry blood smears – thin films. This technique facilitates the extraction of optical anisotropy parameters. Our results establish criteria for distinguishing polycrystalline blood films from those of healthy donors and patients with prostate cancer. Notably, through the integration of polarization-holographic and differential MM methodologies, we introduce, to the best of our knowledge, a novel approach for the spatial 3D characterization of polycrystalline structures within blood films. 

\section{\label{sec:Methods}Methods and Materials}

\subsection{3D Mueller Matrix Imaging Approach}

\begin{figure}[h!]
\begin{center}
\begin{tabular}{c}
\includegraphics[width=1\columnwidth]{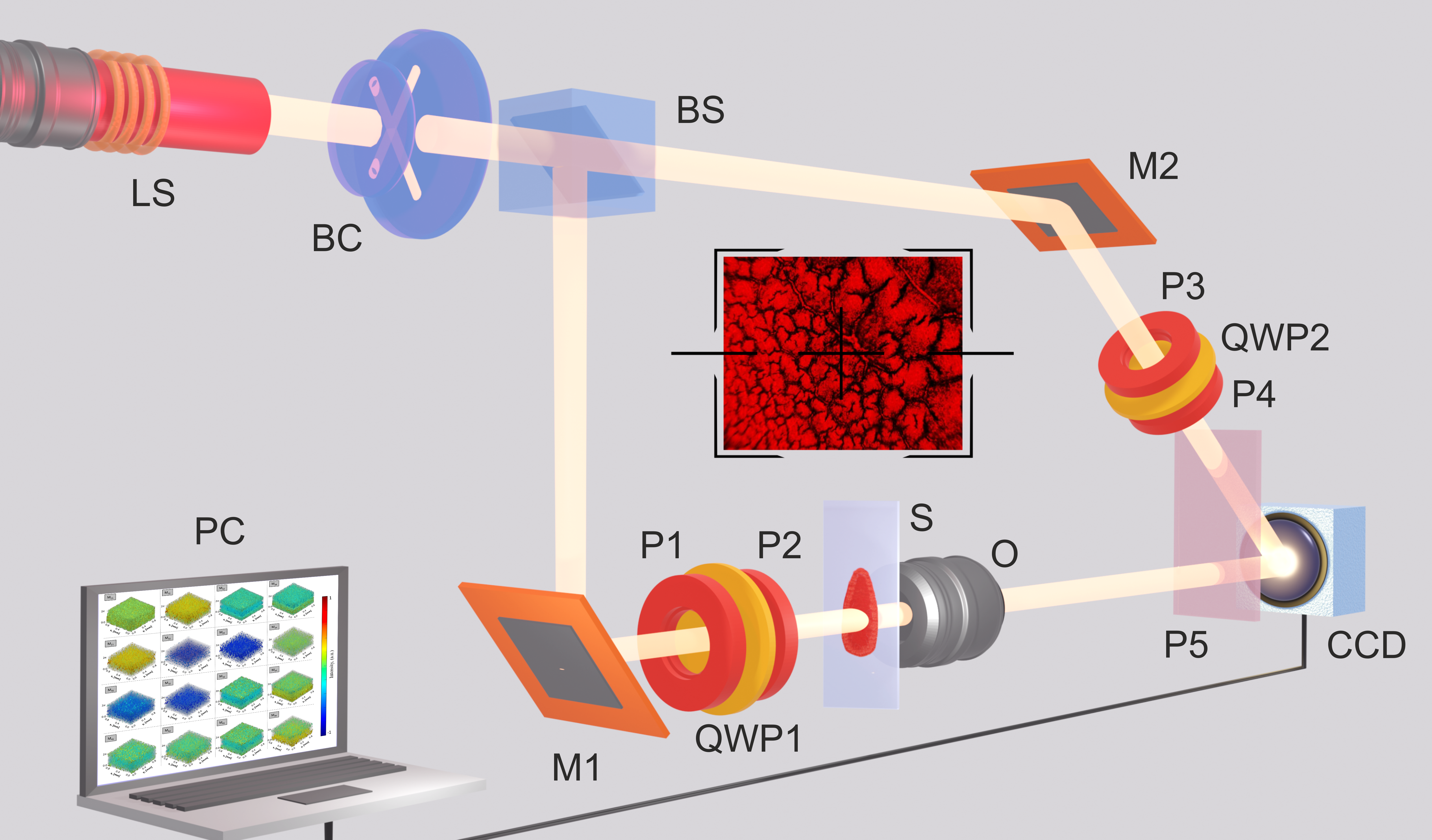}
\end{tabular}
\end{center}
\caption 
{\label{fig:1}
The optical scheme used for MM imaging approach: LS – He-Ne laser; BC – collimator; BS – 50-50 beam splitter; M1,M2 – rotating mirrors;  P1,P2,P3,P4 - linear polarisers; QWP1, QWP2 - quarter wave plate ; S – the polycrystalline blood film sample under investigation; O – objective; P5 – linear polariser (analyser); CCD – digital camera; PC- personal computer. See also \href{https://zenodo.org/doi/10.5281/zenodo.10518536}{\textcolor{cyan}{Supplementary video 1}}} 
\end{figure}

Figure~\ref{fig:1} shows a schematic of the 3D MM imaging experimental setup used in the studies of blood films (see also \href{https://zenodo.org/doi/10.5281/zenodo.10518536}{\textcolor{cyan}{Supplementary video 1}}).
%~\cite{ushenko20213d,ushenko20183d,peyvasteh20203d,ushenko2021embossed}.

A parallel ($2000~\mu m$ in diameter) beam of He-Ne ($633~nm$) laser is collimated before being split into equal "irradiating" and ``reference" beams. Each beam passed through an equivalent polarising filter set to control the polarisation. The irradiating beam passed through the sample, and the image is projected by an objective, through a polariser, into the imaging plane of the digital camera. The reference beam is also guided into the imaging plane of the camera, and an interference pattern is formed from the superposition of the two beams. The camera records the intensity distribution of the interference pattern, which is then computationally analysed.

The layer-by-layer 3D MM polarimetry setup was calibrated using model birefringent phase-shifting plates, including $1/4\lambda$ and $1/2\lambda$ configurations. The accuracy of measuring the magnitude of the MM elements is: $R_{i,j} \sim 0.005$ for $i=1;2;3;j=1;2;3$, and $R_{i,j} \sim 0.01$ for $i=1;2;3;4;j=4;$ and $i=4;j=1;2;3;4$.

The showcased setup functions as a laboratory prototype of a 3D MM tomography tailored for layer-by-layer imaging of the polycrystalline structure in biological tissues and fluids from human organs. Future enhancements are targeted at automating the optical and polarization elements, fine-tuning reconstruction algorithms, and obtaining 3D distributions of anisotropy parameters with the ultimate goal to amalgamate the principles of 3D MM reconstruction with fiber-optic systems, thereby extending the methodology for measuring optical anisotropy parameters \textit{in vivo}.

The presented experimental setup serves as a laboratory prototype for a 3D MM tomography designed specifically for the layer-by-layer imaging of the polycrystalline structure in biological tissues and fluids from human organs~\cite{ushenko2021embossed,ushenko20213d}. Future developments are aimed at automating both the optical and polarization components, refining the reconstruction algorithms, and achieving 3D distributions of anisotropy parameters with the ultimate objective to extend the methodology for robotic automatic standalone optical biopsy and definitive histopathology diagnostics.

\subsection{Samples of blood films}

\begin{figure}[h!]
\begin{center}
\includegraphics[width=1\columnwidth]{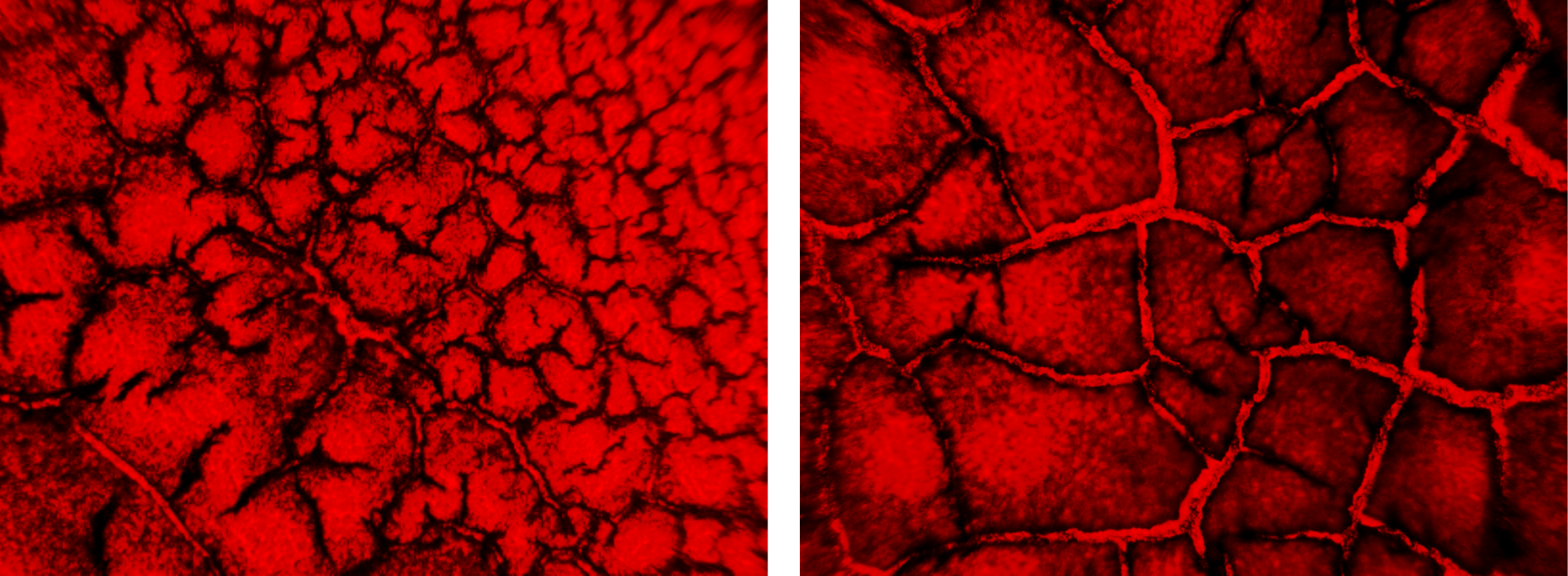}
\end{center}
\caption 
{\label{fig:2}Microscopic images of a structure of representative blood film samples normal (left) and abnormal (right) used in the study; $300 \times 200~\mu m$ each.} 
\end{figure}

In the current study, blood smears are considered as a primary example of evaporated biological liquids. These tiny ($2-5~\mu m$) blood films exhibit a heterogeneous, complex 3D polycrystalline structure (Fig.~\ref{fig:2}) characterized by varying light scattering multiplicities and distinctive depolarizing capabilities. 

In essence, a biological fluid film represents a spatially inhomogeneous and optically anisotropic structure, comprised of various biochemical and molecular crystalline complexes. This film contains elements characterized by multiple optical scattering, including erythrocytes, platelets, and leukocytes, which exhibit circular birefringence and dichroism~\cite{brutin2011pattern}.

For the experiment, polycrystalline blood film samples were collected from both healthy and diseased volunteers. The blood film samples were prepared by applying a blood drop to an optically homogeneous cover glass heated up to $36.6^{\circ}$ in advance. The blood drops fully dehydrated within 40-45 minutes.

Three distinct groups of blood film samples were created:

Group 1 included $k=36$ samples from healthy volunteers.

Group 2 included $k=36$ samples from volunteers diagnosed with moderately differentiated prostate adenocarcinoma ($3+3$ on Gleason’s pattern scale).

Group 3 included $k=36$ samples from volunteers diagnosed with poorly differentiated prostate adenocarcinoma ($4+4$ on Gleason’s Pattern scale).

Table~\ref{tab:table-2}. Optical properties of polycrystalline blood film samples for the groups.

\begin{table}[!ht]
\centering
\caption{Optical parameters of polycrystalline blood film samples}
\label{tab:table-2}
\begin{tabular}{cccc}
\hline
   Parameter     &    Group 1     &  Group 2     & Group 3  \\ \hline
Attenuation (extinction)        &    $0.64 \pm $
  & $0.66 \pm $ & $0.62 \pm $   \\ 
 coefficient $\tau, cm^{-1}$ &  $\pm 0.035$
  & $ \pm 0.039$ & $ \pm 0.032$
  \\ \hline
  Depolarisation degree      &    $39 \pm $
  & $44 \pm $ & $42 \pm $ \\
  $\Lambda,\% $ &    $ \pm 0.77$
  & $\pm 0.084$ & $ \pm 0.81$ 
 \\ \hline
\end{tabular}
\end{table}

The extinction coefficient ($\tau, \text{cm}^{-1}$) of polycrystalline blood film samples is determined following the established photometric method, measuring the attenuation of illuminating beam intensity by the sample~\cite{marchesini1989extinction}. This process utilized an integral light-scattering sphere~\cite{edwards1961integrating}. Additionally, the integral degree of depolarization ($\Lambda, \%$) for polycrystalline blood film samples is assessed utilizing standard MM polarimetry~\cite{swami2013conversion,tuchin2015tissue,izotova1997investigation}. 

To determine the statistical significance of a representative sampling of the number of samples by the cross-validation method~\cite{goodman1975statistical}, the standard deviation $\sigma^2$ of each of the calculated values of the statistical moments $Z_{i=1;2;3;4}(k)$ is determined. The specified number (36 for each group) of samples provided the level $\sigma^2\leq0.025$. This standard deviation corresponds to a confidence interval $p<0.05$, demonstrating the statistical reliability of the 3D MM mapping method.

The sample preparation procedure adhered to the principles of the Declaration of Helsinki and complied with the International Conference on Harmonization-Good Clinical Practice and local regulatory requirements. The study received review and approval from the appropriate Independent Ethics Committees, and written informed consent is obtained from all subjects prior to study initiation.

\begin{table*}[!ht]
\centering
\caption{Structural and logical scheme of the 3D layer-by-layer MM image reconstruction approach. } 
\label{tab:table-1}
\begin{tabular}{cccc}
\hline
   \textnumero      &    Task      &  Method      & Result  \\ \hline
1     &    Extraction of direct  & Differential MM & Algorithms for the reconstruction of  \\
 & information about the & Mapping & the values of linear and circular \\
 & distribution of optical & & birefringence and dichroism \\
 & anisotropy parameters & & averaged over the volume of a \\  
& & & polycrystalline layer \\ \hline
2    &  Reducing the influence of  & Polarization- & Algorithms for layer-by-layer \\
& the depolarized & holographic recording & restoration of the amplitude-phase \\
& background & and restoration of the & structure of the object field of a \\
& & object field & polycrystalline layer \\ \hline
3    & Obtaining layer-by-layer & Phase scanning of the  & Layer- by- layer coordinate  \\
& distributions of optical & amplitude-phase & distributions of linear and circular \\
& anisotropy parameters of a & structure of the & birefringence and dichroism values \\
& polycrystalline layer & object field & \\ \hline
4 & \multicolumn{3}{c}{Synthesis of methods 1-3 for the polarization-holographic investigation of polycrystalline }\\
 & \multicolumn{3}{c}{structure of blood films} \\ \hline
 5 & \multicolumn{3}{c}{Statistical analysis of layer-by-layer maps of linear and circular birefringence and dichroism }\\
 & \multicolumn{3}{c}{of polycrystalline structure of blood films} \\ \hline
 6 & \multicolumn{3}{c}{Data analysis and determination of the diagnostic power (sensitivity, specificity, accuracy)  }\\
 & \multicolumn{3}{c}{of early diagnosis of prostate cancer} \\ \hline
\end{tabular}
\end{table*}

\subsection{3D Mueller Matrix Imaging Approach: Basic equations and theoretical remarks} 

Traditionally, samples containing spatially inhomogeneous optically anisotropic diffuse layers are studied using MM polarimetry approaches~\cite{manhas2006mueller,deng2007characterization,guo2013study,lu1996interpretation,deboo2004degree,buscemi2013near,manhas2015demonstration,pierangelo2013multispectral}. In this approach, only indirect and averaged data (in the form of 16 matrix elements) can be obtained, representing the entire volume of scattering (depolarizing) inhomogeneities. To develop a new, more sensitive, and unambiguous method for tissue diagnosis, it is necessary to comprehensively address several theoretical and experimental challenges and synthesize the obtained results. The principles and steps of the proposed research are outlined in Table~\ref{tab:table-1}.

\subsubsection{Differential Mueller Matrix Mapping }

When photons traverse a depolarizing medium and experience multiple scattering events, alterations occur in the MM of the depolarizing layer along the direction of light propagation~\cite{ortega2011depolarizing,ortega2011mueller,ossikovski2014statistical,ossikovski2014general,devlaminck2013physical,devlaminck2014uniqueness}. Analytically, this relationship is described as: 
\begin{equation}
d\frac{\{R\}\{z\}}{dz}=\{R\}\{z\}\{W\}(z),
\label{eq:1}
\end{equation}
where $\{R\}(z)$ is the MM of an object in a plane at $z$ in the direction of propagation, and $\{W\}(z)$ is the corresponding differential MM. For optically thin, non-depolarizing layers, the differential matrix ${W}(z)$ incorporates six elementary polarization properties, collectively providing a complete characterization of the optical anisotropy of the biological layer
\begin{equation}
\big\{W\big\}=\left\|\begin{array}{cccc}
0 & LD & LD' & CD \\
LD & 0 & CB & -LB' \\
LD' & -CB & 0 & LB \\
CD & LB' & -LB & 0
\end{array}\right\|.
\label{eq:2}
\end{equation}
Here, $LD$ and $LB$ are the linear dichroism and birefringence for a direction of the optical axis $y=0^{\circ}$; $LD'$ and $LB'$ are the linear dichroism and birefringence for a direction of the optical axis $y=45^{\circ}$; and $CD$ and $CB$ are the circular dichroism and birefringence. For a diffuse medium, the matrix~\ref{eq:2} can be represented as separate average $\textlangle\{W\}\textrangle$ (polarisation part $\{W\}(z)$) and fluctuating $\textlangle\{\tilde{W}\}\textrangle$ (depolarisation part $\{\tilde{W}\}(z)$) components
\begin{equation}
\{W\}(z)=\textlangle\{W\}\textrangle(z)+\{\tilde{W}\}(z).
\label{eq:3}
\end{equation}
While the connections between the fluctuating component ${\tilde{W}}(z)$ of the differential matrix (~\ref{eq:3}) and the depolarization parameters of the scattered radiation have been employed in prior diagnostic studies~\cite{bachinsky2021multi,ossikovski2014general}, our focus here is on exploring the potential for reproducing the polarization component $\textlangle\{\tilde{W}\}\textrangle(z)$ of the diffuse biological layer. ased on the preceding theoretical analysis~\cite{buscemi2013near,manhas2015demonstration,pierangelo2013multispectral,ushenko2013spatial,ushenko2013spatiala}, we derive a conclusive expression for the matrix operator
\begin{equation}
 \begin{split}
&\big\{W\big\}=\left\|\begin{array}{cccc}
0 & \textlangle LD\textrangle & \textlangle LD'\textrangle & \textlangle CD \textrangle \\
\textlangle LD\textrangle & 0 & \textlangle CB\textrangle & \textlangle-LB'\textrangle \\
\textlangle LD'\textrangle & \textlangle-CB\textrangle & 0 & \textlangle LB\textrangle \\
\textlangle CD\textrangle & \textlangle LB'\textrangle & \textlangle-LB\textrangle & 0
\end{array}\right\| = \\
& = \left\|\begin{array}{cccc}
0 & ln(r_{12}r_{21}) & ln(r_{13}r_{31})  & ln(r_{14}r_{41}) \\
ln(r_{12}r_{21})  & 0 & ln(\frac{r_{23}}{r_{32}}) & ln(\frac{r_{24}}{r_{42}}) \\
ln(r_{13}r_{31}) & ln(\frac{r_{32}}{r_{23}}) & 0 & ln(\frac{r_{34}}{r_{43}}) \\
ln(r_{14}r_{41}) & ln(\frac{r_{42}}{r_{24}})& ln(\frac{r_{43}}{r_{34}}) & 0
\end{array}\right\|.
\end{split}
\label{eq:4}
\end{equation}
By collectively examining equations (\ref{eq:2}) and (\ref{eq:4}), algorithms can be deduced for reproducing the average values of the phase and amplitude anisotropy parameters:
\begin{equation}
\textlangle\ LB\textrangle(\delta)=\frac{2\pi z}{\lambda}\Delta n_{LB}=ln(\frac{r_{34}}{r_{42}});
\label{eq:5}
\end{equation}
\begin{equation}
\textlangle\ LB'\textrangle(\delta^*)=\frac{2\pi z}{\lambda}\Delta n^{*}_{LB}=ln(\frac{r_{24}}{r_{42}});
\label{eq:6}
\end{equation}
\begin{equation}
\textlangle\ CB\textrangle(\phi)=\frac{2\pi z}{\lambda}\Delta n^{*}_{CB}=ln(\frac{r_{23}}{r_{32}});
\label{eq:7}
\end{equation}
\begin{equation}
\textlangle\ LD\textrangle(\Delta\tau_{(0^{\circ}-90^{\circ})})=tgy=ln(r_{12}r_{21});
\label{eq:8}
\end{equation}
\begin{equation}
\textlangle\ LD'\textrangle(\Delta\tau_{(45^{\circ}-135^{\circ})})=tg(y+45^{\circ})=ln(r_{13}r_{31});
\label{eq:9}
\end{equation}
\begin{equation}
\textlangle\ CD\textrangle(\Delta g)=\frac{\chi_{\otimes}-\chi_{\oplus}}{\chi_{\otimes}+\chi_{\oplus}}=ln(r_{14}r_{41});
\label{eq:10}
\end{equation}
where $\delta$ and $\delta^*$ are the phase shifts between orthogonally polarised ($0^{\circ}-90^{\circ}$ and $45^{\circ}-135^{\circ}$) components of the amplitude of incident light; $\Delta n_{LB}$ and $\Delta n_{LB}^{*}$ are the magnitudes of the linear birefringence for $0^{\circ}-90^{\circ}$ and $45^{\circ}-135^{\circ}$ respectively; $\phi$ is the phase shift between the right- ($\otimes$) and left- ($\oplus$) circularly polarised components of the  amplitude of laser radiation; $\Delta n_{CB}$ is the circular birefringence; $\Delta\tau_{(0^{\circ}-90^{\circ})}$and $\Delta\tau_{(45^{\circ}-135^{\circ})}$ are the ratios between the absorption coefficients of orthogonally polarised ($0^{\circ}-90^{\circ}$ and $45^{\circ}-135^{\circ}$) components of the amplitude of laser radiation; $\chi_{\otimes}$ and $\chi_{\oplus}$ are,  respectively, the absorption coefficients of the right and left circularly polarised components of the laser radiation amplitude; and $\lambda$ is the laser wavelength.

\subsubsection{Polarization-holographic recording and restoration of the object field }

To determine the layer-by-layer distributions of matrix elements $r_{ik}$ six distinct polarization states are formed in the illuminating ($Ir$) and reference ($Re$) beams $(\{Ir-Re\}\Rightarrow 0^{\circ}; 90^{\circ}; 45^{\circ}; 135^{\circ}; \otimes; \oplus$). For each polarization state ($p$ $i$ $r$), two partial interference patterns are registered through a polarizer-analyzer oriented at $\Omega=0^{\circ};\Omega=90^{\circ}$. For each partial interference distribution, two-dimensional discrete Fourier transform $F(u,v)$ is further performed. The $F(u,v)$ of a two-dimensional array $I_{\Omega=0^{\circ};90^{\circ})}(m,n)$ (the obtained image) is a function of two discrete variables coordinates $(m,n)$ camera pixels defined by~\cite{ushenko2021embossed}:
\begin{equation}
\begin{split}
\displaystyle
&\Phi F_{x;y}(\Omega=0^{\circ};90^{\circ})(u,v)=\\
&=\frac{1}{M\times N}\sum^{M-1}_{m=0}\sum^{N-1}_{n=0} I_{x,y}(\Omega=0^{\circ};90^{\circ})(m,n)\times\\
&\times\Bigl[-i2\pi\Bigl(\frac{m\times u}{M}+\frac{n\times v}{N}\Bigr) \Bigr],
\label{eq:11}
\end{split}
\end{equation}
where 
\begin{equation}
\begin{split}
\left\{\begin{array}{llll}
&I_{x}^{0}(\Omega=0^{\circ})(m,n)=\\
&=U_{x}^{0}(\Omega=0^{\circ})(m,n)(U_{x}^{0})^{*}(\Omega=0^{\circ})(m,n);
\\
&I_{x}^{90}(\Omega=90^{\circ})(m,n)=\\
&=U_{x}^{90}(\Omega=90^{\circ})(m,n)(U_{x}^{90})^{*}(\Omega=90^{\circ})(m,n);
\\
&I_{y}^{0}(\Omega=0^{\circ})(m,n)=\\
&=U_{y}^{0}(\Omega=0^{\circ})(m,n)(U_{y}^{0})^{*}(\Omega=0^{\circ})(m,n);
\\
&I_{y}^{90}(\Omega=90^{\circ})(m,n)=\\
&=U_{y}^{90}(\Omega=90^{\circ})(m,n)(U_{y}^{90})^{*}(\Omega=90^{\circ})(m,n);
\end{array}\right .
\label{eq:12}
\end{split}
\end{equation}
are the coordinate distributions of the intensity of the interference pattern filtered by the analyser with the orientation of its transmission axis at $\Omega=0^{\circ};\Omega=90^{\circ};$ $^*$ denotes the complex conjugation operation; ($u,v$) are the spatial frequencies in the x and y directions respectively; and ($M,N$) are the number of pixels of the CCD camera in the m and n directions respectively, such that $0\leq m$,$u \leq M$ and $0\leq n$,$v \leq N$. The subsequent application of the two-dimensional inverse discrete Fourier transform on the obtained spectrum can be expressed as
\begin{equation}
\begin{split}
\displaystyle
&\bigl[(\Phi T^{*}) \bigr]_{\Omega=0^{\circ};90^{\circ}}^{0^{\circ};90^{\circ};45^{\circ};135^{\circ};\otimes;\oplus}(u,v)=
\\
&=  \frac{1}{M\times N}\sum_{m=0}^{M-1}\sum_{n=0}^{N-1}I_{\Omega=0^{\circ};90^{\circ}}^{0^{\circ};90^{\circ};45^{\circ};135^{\circ};\otimes;\oplus}(m,n)\times\\
&\Bigl[-i2\pi\Bigl(\frac{m\times u}{M}+\frac{n\times v}{N}\Bigr) \Bigr].
\end{split}
\label{eq:13}
\end{equation}
Here,
\begin{equation}
\begin{split}
&\bigl[(\Phi T^{*}) \bigr]_{\Omega=0^{\circ};90^{\circ}}^{0^{\circ};90^{\circ};45^{\circ};135^{\circ};\otimes;\oplus}(m,n)^{*}(x,y)\equiv\\
&\equiv U_{\Omega=0^{\circ};90^{\circ}}^{0^{\circ};90^{\circ};45^{\circ};135^{\circ};\otimes;\oplus}(m,n).
\end{split}
\label{eq:14}
\end{equation}

Ultimately, the complex amplitude distribution for each polarization state can be derived in various phase planes $\theta_{k}=(\delta_{y}-\delta_{x})$ of the object field, separated by an arbitrary step of $\Delta\theta$:
\begin{equation}
\left\{\begin{array}{lrlr}
\Omega_{0^{\circ}}\rightarrow|U_{x}^{0}(\Omega=0^{\circ})|;
\\
\Omega_{90^{\circ}}\rightarrow|U_{x}^{90}(\Omega=90^{\circ})|exp(i(\delta_{x}^{90}-\delta_{x}^{0})),
\end{array}\right .
\label{eq:15}
\end{equation}
\begin{equation}
\left\{\begin{array}{lrlr}
\Omega_{0^{\circ}}\rightarrow|U_{y}^{0}(\Omega=0^{\circ})|;
\\
\Omega_{90^{\circ}}\rightarrow|U_{y}^{90}(\Omega=90^{\circ})|exp(i(\delta_{y}^{90}-\delta_{y}^{0})),
\end{array}\right .
\label{eq:16}
\end{equation}

\subsubsection{Phase scanning of the amplitude-phase structure of the object field}

The algorithm, described by (\ref{eq:15}) and (\ref{eq:16}), for scanning the phase of the complex amplitude field (\ref{eq:13}) and (\ref{eq:14}) directly corresponds to the physical depth $h_i$ of an optically anisotropic biological layer in the case of single scattering:
\begin{equation}
 h_i=\frac{\lambda}{2\pi\Delta n}\theta_i.
\label{eq:17}
\end{equation}

In the case of multiple scattering, the physical depth is multiplied (effective optical depth $h_{i}^{*}$) and becomes a multiple of the geometric thickness value of the biological layer $z$.

\begin{equation}
 h_{i}^{*}\sim Kz.
\label{eq:18}
\end{equation}

In each phase plane $\theta_k$ the corresponding sets of parameters of the Stokes vector and polarization parameters of the object field of the biological layer can be calculated as:

\begin{equation}
\begin{split}
 \left\{\begin{array}{llll}
&ST_{1}^{0^{\circ};90^{\circ};45^{\circ};135^{\circ};\otimes;\oplus}(\theta_k,m,n)=\\
&=(|U_0|^2+|U_{90}|^2)(\theta_k,m,n);
\\
&ST_{2}^{0^{\circ};90^{\circ};45^{\circ};135^{\circ};\otimes;\oplus}(\theta_k,m,n)=\\
&=(|U_0|^2-|U_{90}|^2)(\theta_k,m,n);
\\
&ST_{3}^{0^{\circ};90^{\circ};45^{\circ};135^{\circ};\otimes;\oplus}(\theta_k,m,n)=\\
&=2Re(U_{0}U_{90}^*)(\theta_k,m,n);
\\
&ST_{3}^{0^{\circ};90^{\circ};45^{\circ};135^{\circ};\otimes;\oplus}(\theta_k,m,n)=\\
&2Im(U_{0}U_{90}^*)(\theta_k,m,n).
\end{array}\right .
\end{split}
\label{eq:19}
\end{equation}
Based on relations~\ref{eq:19}, the set of elements of the MM $\{R\}$ is calculated using the following Stokes-polarimetric relations:

\begin{equation}
\resizebox{1\hsize}{!}{$
 \begin{split}
&\big\{R\big\}(\theta_k,m,n)=\left\|\begin{array}{cccc}
r_{11} & r_{12} & r_{13}& r_{14} \\
r_{21} & r_{22} & r_{23} & r_{24} \\
r_{31} &r_{32} & r_{33} & r_{34} \\
r_{41} & r_{42} & r_{43} & r_{44}
\end{array}\right\| =0.5(\theta_k,m,n)\times \\
& \times \left\|\begin{array}{cccc}
(ST_{1}^{0}+ST_{1}^{90}) & (ST_{1}^{0}-ST_{1}^{90}) & (ST_{1}^{45}-ST_{1}^{135})  & (ST_{1}^{\otimes}-ST_{1}^{\oplus}) \\
(ST_{2}^{0}+ST_{2}^{90}) & (ST_{2}^{0}-ST_{2}^{90}) & (ST_{2}^{45}-ST_{2}^{135})  & (ST_{2}^{\otimes}-ST_{2}^{\oplus}) \\
(ST_{3}^{0}+ST_{3}^{90}) & (ST_{3}^{0}-ST_{3}^{90}) & (ST_{3}^{45}-ST_{3}^{135})  & (ST_{3}^{\otimes}-ST_{3}^{\oplus}) \\
(ST_{4}^{0}+ST_{4}^{90}) & (ST_{4}^{0}-ST_{4}^{90}) & (ST_{4}^{45}-ST_{4}^{135})  & (ST_{4}^{\otimes}-ST_{4}^{\oplus})
\end{array}\right\|.
\end{split}
$}
\label{eq:20}
\end{equation}

Using the set of distributions ~\ref{eq:4}-~\ref{eq:10} a series of layer-by-layer distributions of the mean values of linear and circular birefringence and dichroism ($G(\textlangle LB \textrangle,\textlangle LB' \textrangle, \textlangle CB \textrangle, \textlangle LD \, \textlangle LD' \textrangle, \textlangle CD \textrangle)$) can be obtained :
\begin{equation}
\textlangle\ LB\textrangle(\theta_k,m,n)=ln\Bigl(\frac{(ST_{3}^{\otimes}-ST_{3}^{\oplus})}{(ST_{4}^{45}-ST_{4}^{135})}\Bigr)(\theta_k,m,n);
\label{eq:21}
\end{equation}
\begin{equation}
\textlangle\ LB'\textrangle(\theta_k,m,n)=ln\Bigl(\frac{(ST_{2}^{\otimes}-ST_{2}^{\oplus})}{(ST_{4}^{0}-ST_{4}^{90})}\Bigr)(\theta_k,m,n);
\label{eq:22}
\end{equation}
\begin{equation}
\textlangle\ CB\textrangle(\theta_k,m,n)=ln\Bigl(\frac{(ST_{2}^{45}-ST_{2}^{135})}{(ST_{3}^{0}-ST_{3}^{90})}\Bigr)(\theta_k,m,n);
\label{eq:23}
\end{equation}
\begin{equation}
\begin{split}
&\textlangle\ LD\textrangle(\theta_k,m,n)=\\
&=ln((ST_{1}^{0}-ST_{1}^{90})(ST_{2}^{0}+ST_{2}^{90}))(\theta_k,m,n);
 \end{split}
\label{eq:24}
\end{equation}

\begin{equation}
\begin{split}
&\textlangle\ LD'\textrangle(\theta_k,m,n)=tg(y+45^{\circ})=\\
&=ln((ST_{1}^{45}-ST_{1}^{135})(ST_{3}^{0}+ST_{3}^{90}))(\theta_k,m,n);
\end{split}
\label{eq:25}
\end{equation}

\begin{equation}
\begin{split}
&\textlangle\ CD\textrangle(\theta_k,m,n)=\\
&=ln((ST_{1}^{\otimes}-ST_{1}^{\oplus})(ST_{4}^{0}+ST_{4}^{90}))(\theta_k,m,n);
\end{split}
\label{eq:26}
\end{equation}
According to~\cite{ortega2011depolarizing,ortega2011mueller,ossikovski2014statistical,ossikovski2014general,devlaminck2013physical,devlaminck2014uniqueness}, we will further operate with the generalized quantities of linear birefringence and dichroism:
\begin{equation}
LB\equiv\sqrt{\textlangle\ CB\textrangle^2+\textlangle\ CB'\textrangle^2};
\label{eq:27}
\end{equation}
\begin{equation}
LD\equiv\sqrt{\textlangle\ LD\textrangle^2+\textlangle\ CD\textrangle^2};
\label{eq:28}
\end{equation}
In this manner, the synthesis of the 1st order differential matrix (\ref{eq:1})--(\ref{eq:10}) and the algorithms for layer-by-layer polarization-holographic determination of matrix elements (\ref{eq:11})--(\ref{eq:20}) allows the acquisition of layer-by-layer maps of linear and circular birefringence and dichroism of the polycrystalline structure of the dehydrated blood films (\ref{eq:21})--(\ref{eq:28}).

\subsubsection{Quantitative Evaluation of Polarization Maps}

To assess the layer-by-layer maps of optical anisotropy ($G$), statistical moments of the first ($Z_1$), second ($Z_2$), third ($Z_3$), and fourth ($Z_4$) orders are utilized and calculated as follows~\cite{swami2013conversion,tuchin2015tissue,izotova1997investigation}:
\begin{eqnarray*}
\begin{array}{lrlr}
\displaystyle
Z_1 = {\frac{1}{P}} \sum_{j=1}^{P} (G(\theta,m\times n))_j;\\
\displaystyle
Z_2=\sqrt{\frac{1}{P}\sum_{j=1}^{P} (G^2(\theta,m\times n))_j};\\
\displaystyle
Z_3=\frac{1}{Z_2^3}\frac{1}{P}\sum_{j=1}^{P} (G^3(\theta,m\times n))_j;\\
\displaystyle
Z_4=\frac{1}{Z_2^4}\frac{1}{P}\sum_{j=1}^{P} (G^43(\theta,m\times n))_j,\\
\end{array}
\label{eq:29}
\end{eqnarray*}
where $P=m\times n$ is the $x–y$ resolution of the camera. These measures ($Z_1$--$Z_4$) most fully characterize the mean, variance, skewness, and kurtosis of the layer-by-layer ($\theta_k$) distributions $G(\theta_k,m,n)$ of linear and circular birefringence ($LB, \langle CB \rangle$) and dichroism ($LD, \langle CD \rangle$).

The methodology for implementing this statistical approach involves several steps. Initially, groups of blood polycrystalline film samples are formed from both healthy and diseased patients. For each sample within each group, anisotropy maps, denoted as $G(\theta_k,m,n)$, are obtained for a series ($H$) of phase sections $\theta_{(k=1...H)}$. Subsequently, a set of statistical moments of 1st to 4th orders, denoted as $Z_{i=1-4}$, is calculated using Eq.~\ref{eq:29}. 

\begin{figure*}[ht!]
\begin{center}
\begin{tabular}{c}
\includegraphics[width=1\linewidth]{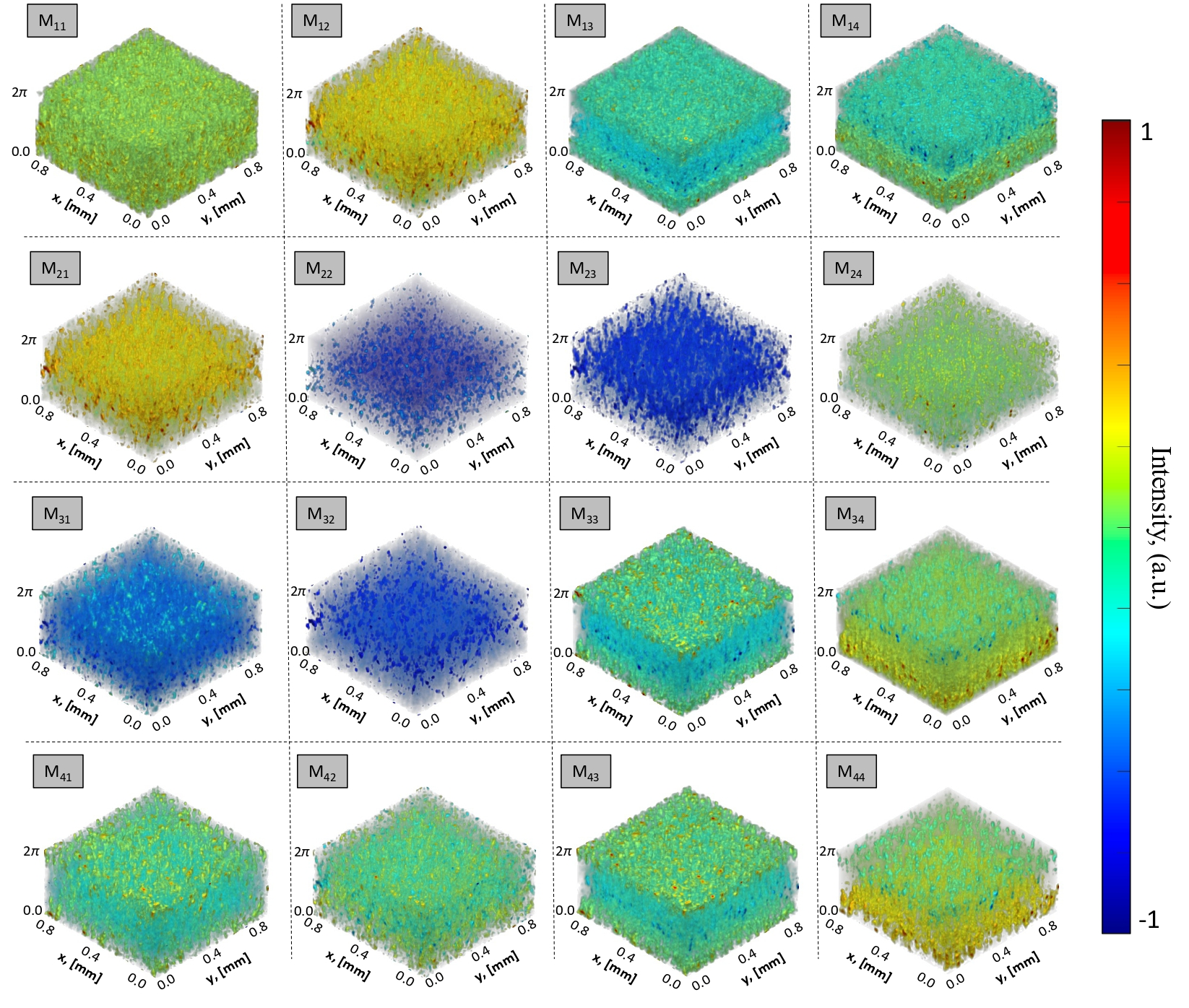}
\end{tabular}
\end{center}
\caption 
{\label{fig:3}
An example of typical polarization-interference measurement through digital holographic reconstruction, showcasing layered 3D elements of the MM for a dehydrated blood film.} 
\end{figure*} 

Within each group, the mean value $\bar{Z}{i=1-4}$ and standard deviation $\sigma{i=1-4}$ are computed for the obtained distribution of $Z_{i=1-4}$. Further, statistically significant ($p\leq0.05$) intergroup differences (``norma" – ``cancer" – ``stage") for each phase section $\theta_{k=1...H}$ are determined for each statistical moment of 1st to 4th orders $Z_{i=1-4}$ ($LB,\textlangle CB \textrangle,LD,\textlangle CD\textrangle$). The measures $Z_{i=1-4}^*$ ($LB,\textlangle CB \textrangle,LD,\textlangle CD\textrangle$) are then employed in algorithms for the information analysis of evidence-based medicine~\cite{cassidy2005basic,davis2002statistical,robinson1996principles}.

\subsubsection{Quantitative Analysis of Anisotropy Maps}

For each statistically significant parameter $Z_{i=1-4}^*$ ($LB,\textlangle CB \textrangle,LD,\textlangle CD\textrangle$) the criteria of evidence-based medicine has been used~\cite{cassidy2005basic,davis2002statistical,robinson1996principles}:

1) Sensitivity ($Se$) - proportion of true positive results ($TP$) among the group of diseased ($D_+$) patients 
\begin{equation}
Se=\frac{TP}{D_+}100\%,
\label{eq:30}
\end{equation}
2) Specificity ($Sp$) - proportion of true negative results ($TN$) among the control group of healthy patients ($D_-$)
\begin{equation}
Sp=\frac{TP}{D_-}100\%,
\label{eq:31}
\end{equation}
3) 	Accuracy ($Ac$) – proportion of true results ($TP+TN$) among all the patients ($D_++D_-$)
\begin{equation}
Ac=\frac{TP+TN}{D_++D_-}100\%.
\label{eq:32}
\end{equation}
In our study, accuracy refers to the quantity of accurate diagnoses achieved through the utilization of 3D layer-by-layer MM reconstruction for anisotropy mapping of the polycrystalline structure in blood films.

Figure \ref{fig:3} illustrates an example of polarization-interference measurement using digital holographic reconstruction of layered 3D elements of the MM of a dehydrated blood film. Each 3D matrix element comprises a set of 200 phase-resolved 2D layers (ranging from $0$ to $2\pi$ with a scanning step of $0.01\pi$). For each phase-resolved layer, the complex amplitude field (\ref{eq:16}) was reconstructed using algorithm (\ref{eq:13}). Subsequently, in this phase plane, the coordinate distributions of the MM elements are computed utilizing (\ref{eq:20}) and (\ref{eq:21}). Thus, through phase scanning of 3D matrix elements, optical anisotropy maps (corresponding to Eqs. (\ref{eq:5})--(\ref{eq:10})) are reconstructed in each phase plane.

In current study, three phase slices at 0.2, 0.6, and 1.0 radians were utilized. The provided example illustrates the presence of all types of optical anisotropy mechanisms in the dehydrated blood film, as evidenced by the asymmetry in experimentally measured matrix elements. This observation can be theoretically explained by considering the presence of a complex mechanism involving linear birefringence and linear dichroism. The optical manifestations of linear birefringence in this mechanism are characterized by the partial MM:
\begin{equation}
\big\{D\big\}=\left\|\begin{array}{cccc}
1 & 0 & 0& 0 \\
0 & d_{22} & d_{23} & d_{24} \\
0 & d_{32} & d_{33} & d_{34} \\
0 & d_{42} & d_{43} & d_{44}
\end{array}\right\|, 
\label{eq:33}
\end{equation}
where
\begin{equation}
\left\{\begin{array}{lrlr}
d_{22}=cos^2 2\rho + sin^2 2\rho \, cos \, \delta,
\\
d_{23}=d_{32}=cos \,2\rho \, sin \,2 \rho (1- cos \, \delta),
\\
d_{33}=sin^2 2\rho + cos^2 2 \rho \, cos \, \delta,
\\
d_{42}=-d_{24}=sin \,2\rho \, sin \, \delta,
\\
d_{34}=-d_{43}=cos \,2\rho \, sin \, \delta,
\\
d_{44}=cos \, \delta.
\end{array}\right .
\label{eq:34}
\end{equation}
Here, $\rho $ is the optical axis direction, determined by the orientation of the polypeptide chain of amino acids; $\delta=\frac{2\pi}{\lambda}\mathrm{\Delta nl}$ is the phase shift between linearly orthogonal polarized components of the laser beam amplitude; $\lambda$ is the wavelength, $\mathrm{\Delta n}$ is the magnitude of birefringence; $l$ represents the geometric thickness of the layer.

\subsubsection{Linear dichroism}
The analytical expression for the partial matrix operator that characterizes linear dichroism in optically anisotropic absorption is as follows:
\begin{equation}
\big\{\Psi\big\}=\left\|\begin{array}{cccc}
1 & \phi_{12} & \phi_{13}& 0 \\
\phi_{21}& \phi_{22} & \phi_{23} & 0 \\
\phi_{31}& \phi_{32} & \phi_{33} & 0 \\
0 & 0 & 0 & \phi_{44}
\end{array}\right\|, 
\label{eq:35}
\end{equation}
where
\begin{equation}
\left\{\begin{array}{lrlr}
\phi_{12}=\phi_{21}=(1-\Delta\tau)cos\, 2\rho ,
\\
\phi_{13}=\phi_{31}=(1-\Delta\tau)sin\, 2\rho ,
\\
\phi_{22}=(1+\Delta\tau)cos^2\, 2\rho+2\sqrt{\Delta\tau} sin^2\, 2\rho,
\\
\phi_{23}=\phi_{32}=(1-\sqrt{\Delta\tau})^2 cos\, 2\rho \,  sin\, 2\rho ,
\\
\phi_{33}=(1+\Delta\tau)sin^2\, 2\rho+2\sqrt{\Delta\tau} cos^2\, 2\rho,
\\
\phi_{44}=2\sqrt{\Delta\tau}.
\end{array}\right .
\label{eq:36}
\end{equation}
Here, $\mathrm{\Delta\tau}=\frac{\tau_x}{\tau_y}$, 
$\bigl\{\begin{array}{lrlr} \tau_x=\tau\,cos\,\rho;\\\tau_y=\tau\,sin\,\rho, \end{array}$ and $\tau_{x}$, $\tau_{y}$
is the coefficients of absorption for linearly polarized orthogonal components of laser radiation amplitude. 

The resulting operator of two optical anisotropy mechanisms:
\begin{equation}
\big\{F\big\}=\left\|\begin{array}{cccc}
1 & 0 & 0& 0 \\
0 & d_{22} & d_{23} & d_{24} \\
0 & d_{32} & d_{33} & d_{34} \\
0 & d_{42} & d_{43} & d_{44}
\end{array}\right\|\times\left\|\begin{array}{cccc}
1 & \phi_{12} & \phi_{13}& 0 \\
\phi_{21}& \phi_{22} & \phi_{23} & 0 \\
\phi_{31}& \phi_{32} & \phi_{33} & 0 \\
0 & 0 & 0 & \phi_{44}
\end{array}\right\|, 
\label{eq:37}
\end{equation}

Observably, the symmetry of the birefringence matrix operator ($d_{23}=d_{32}$; $d_{34}=-d_{43}$;$d_{24}=-d_{42}$) is disrupted.
\begin{equation}
\begin{array}{c}
  f_{34}\neq f_{43};\\
f_{34}=d_{34}\phi_{44};\\
f_{43}=d_{42}\phi_{23}+d_{43}\phi_{33};\\
f_{23}\neq f_{32};\\
f_{23}=d_{22}\phi_{13}+d_{23}\phi_{22};\\
f_{32}=d_{32}\phi_{12}+d_{33}\phi_{32}.
\end{array}
\label{eq:38}
\end{equation}

\subsection{Diagnostic Algorithmic Framework}
An analytical protocol for distinguishing between normal (healthy) and abnormal (e.g., cancerous prostate) tissues is outlined as follows. Initially, the identification of the phase plane, denoted as $\theta^*$, which demonstrates heightened sensitivity to pathological alterations in the optical anisotropy parameters of the polycrystalline blood structure, is undertaken as:

1) An initial ``macro” phase scanning step $\theta_k^{max}=0.05 , \text{rad}$ is selected.

2) The layer-by-layer coordinate distributions $G(\theta_k,m,n)$ are reconstructed for each $\theta_k^{max}$.

3) The statistical moments $Z_{i=1;2;3;4}$ are calculated.

4) The differences between the values of each of the statistical moments are calculated $(\Delta Z_i)k=Z_i(\theta{j+1}^{max})-Z_i(\theta_{j}^{max})$.

5) The phase interval $\Delta\theta^*=\theta_{j+1}^{max}-\theta_{j}^{max}$ within which the monotonic increase in the value of $(\Delta Z_i)k=Z_i(\theta{j+1}^{max})-Z_i(\theta_{j}^{max})$ stopped is determined.

6) Within the limits $\Delta\theta^*$, a new series of values $(\Delta Z_i)k=Z_i(\theta{q+1}^{min})-Z_i(\theta_{q}^{min})$ is calculated with a discrete “micro” phase scanning step $\theta_q^{min}=0.01~\text{rad}$.

7) For each optical anisotropy parameter in $G$, the optimal phase plane $\theta^*$, in which $\Delta Z_i(\theta^*)=max$, is determined.

8) 	In these planes ($\theta^*$ ), the mean $\bar Z_{i=1;2;3;4}^*$ and standard deviations $\sigma(\Delta Z_i^*)$ are determined for comparing between groups 1 and 2, as well as for comparing between group 2 and 3. The sensitivity ($Se$), specificity ($Sp$) and balanced accuracy ($Ac$) are also calculated~\cite{goodman1975statistical,cassidy2005basic,davis2002statistical,robinson1996principles}. 

\section{Results and Discussions}

\subsection{Layer-by-layer Phase and Amplitude Anisotropy Mapping in Polycrystalline Blood Films: Insights from Healthy Donors}

By employing phase scanning techniques (\ref{eq:15}) and (\ref{eq:16}) on the reconstructed object field of complex amplitudes, we extract layer-by-layer coordinate distributions of optical anisotropy parameters (\ref{eq:21})--(\ref{eq:28}) in blood film samples. The selection of phase planes $\theta_i$ in the object field of complex amplitudes, along with their corresponding physical depths $h_i$ (\ref{eq:17}) and effective depths $h_i^*$ (\ref{eq:18}) in our samples, is guided by optical-geometric approximations. Specifically, we consider $\Delta n\sim 10^{-3}$~\cite{tuchin2015tissue,tuchin2016tissue,tuchin2010handbook} and a wavelength of $\lambda=0.63\mu m$.

Utilizing (\ref{eq:9}) and (\ref{eq:17}), we estimate the phase intervals for scattering of various multiplicities in the object plane. A single pass of laser radiation through the polycrystalline blood film corresponds to the value $\theta_1\approx0.7rad\Leftrightarrow z_1\approx70\mu m$. Similarly, double and triple passes correspond to $\theta_2\approx1.4rad\Leftrightarrow z_2^\ast\approx140\mu m$ and $\theta_3\approx2.1rad\Leftrightarrow z_2^\ast\approx210\mu m$, respectively.

In other words, phase shifts $\theta\le0.7rad$ predominantly represent single scattering, while shifts in the range $0.7rad\le\theta\le1.4rad$ indicate low multiplicity scattering. For $\theta\geq1.4rad$, multiple scattering prevails. Performing scanning in the range of phase shifts $0.15rad\le\theta\le0.7rad$ enables a significant reduction in the influence of depolarized background and enhances the signal for $1\mu m\le h\le70\mu m$. It's important to note that this evaluation doesn't account for the scattering multiplicity of optically active shaped elements at all depths in the polycrystalline blood film. Thus, we choose three phase planes corresponding to three regimes of laser light interaction with inhomogeneities in the blood film:

1) 	$\theta=0.2~rad$ – Characterized by single scattering at both fibrillar networks of proteins (albumin, elastin, fibrin) and optically active shaped elements (erythrocytes, monocytes, leukocytes) in blood.

2) $\theta=0.6~rad$ – Predominantly single scattering at fibrillar networks of proteins with an increased scattering multiplicity at optically active shaped elements in blood.

3) $\theta=1.0~rad$ – Mainly associated with multiple scattering at optically active shaped elements in blood.

\begin{figure*}[ht!]
\begin{center}
\begin{tabular}{c}
\includegraphics[width=2\columnwidth]{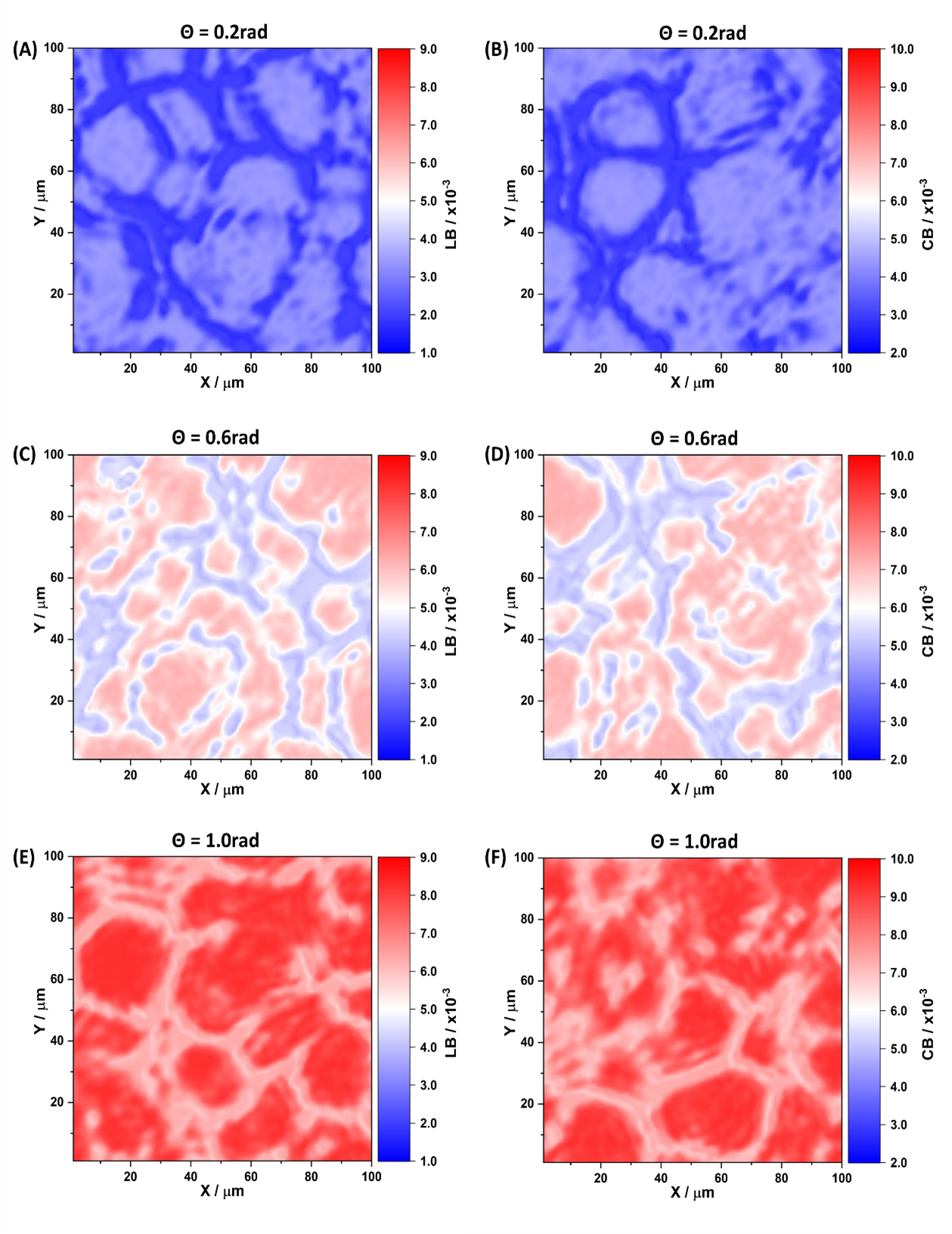}
\end{tabular}
\end{center}
\caption 
{\label{fig:4}
Maps of the linear $LB\left(\theta_k,m,n\right)$ (A,C,E) and circular $CB\left(\theta_k,m,n\right)$ birefringence (B,D,F) of a polycrystalline film of the blood of a healthy donor at ``phase” sections of $0.2rad$ (A-B), $0.6~rad$ (C-D) and $1.0~rad$ (E-F).  } 
\end{figure*}

\begin{figure*}[ht!]
\begin{center}
\begin{tabular}{c}
\includegraphics[width=1.8\columnwidth]{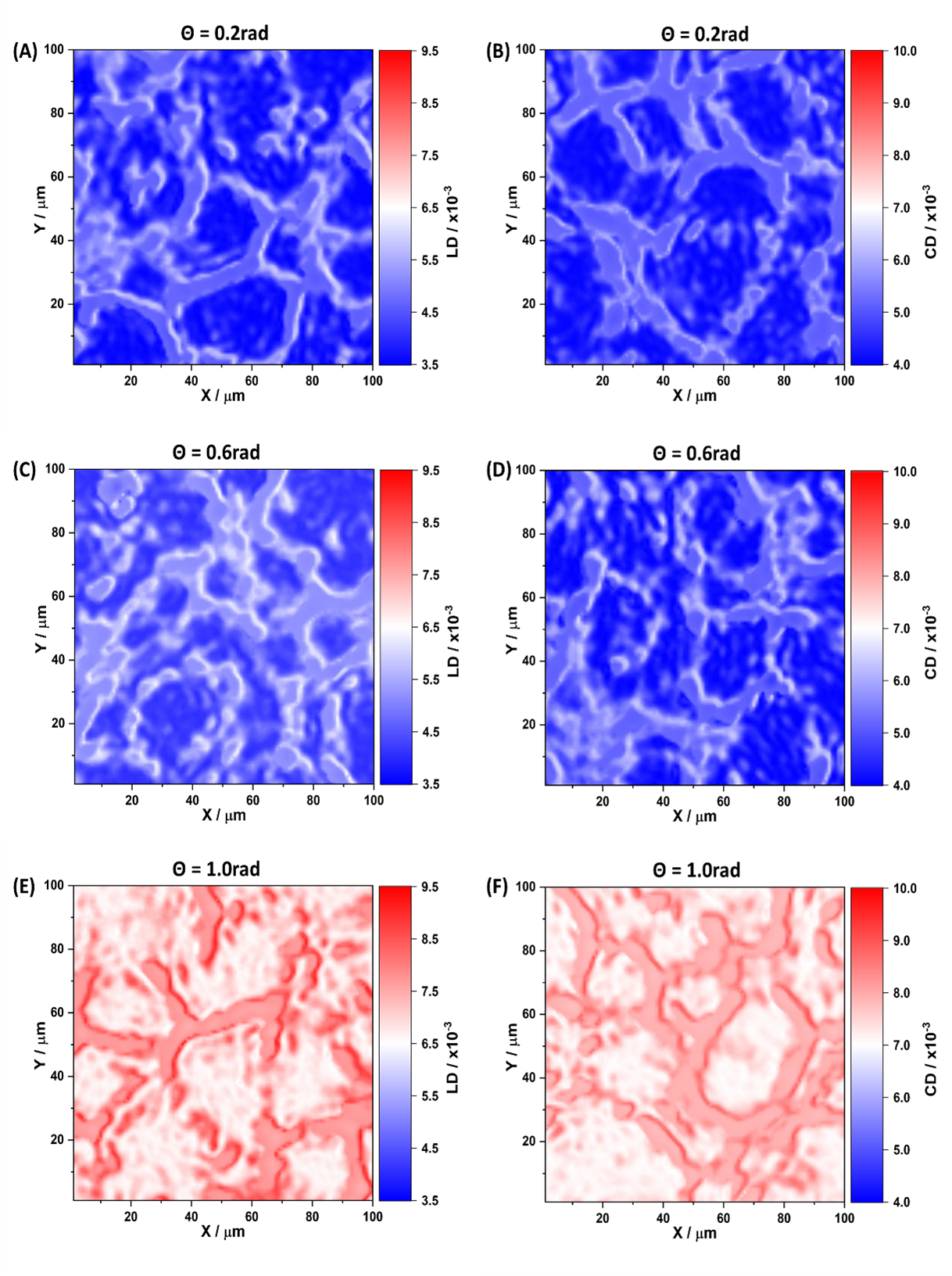}
\end{tabular}
\end{center}
\caption 
{\label{fig:5}
Maps of the linear $LD\left(\theta_k,m,n\right)$ (A,C,E) and circular $CD\left(\theta_k,m,n\right)$ dichroism (B,D,F) of a polycrystalline film of the blood of a healthy donor at ``phase” sections of $0.2~rad$ (A-B), $0.6~rad$ (C-D) and $1.0~rad$ (E-F). } 
\end{figure*}

Figure~\ref{fig:4} and Figure~\ref{fig:5} depict maps illustrating the phase and amplitude anisotropies of the blood film for a series of phase planes $\theta_k=0.2rad;0.6rad;1.0rad$. The analysis of the layer-by-layer maps of the phase (Figure~\ref{fig:4}) and amplitude (Figure~\ref{fig:5}) anisotropies of the polycrystalline blood film reveals several key findings. Firstly, all types of optical anisotropy, denoted as $G$ and including linear birefringence ($LB$), circular birefringence ($\left\langle CB\right\rangle$), linear dichroism ($LD$), and circular dichroism ($\left\langle CD\right\rangle$), are present in the polycrystalline structure of the blood film. This observation indicates the existence of supramolecular structural anisotropy, specifically $LB$ and $LD$, formed by the polycrystalline networks of protein molecules. It also suggests that optically active shaped elements of blood contribute to the formation of circular birefringence ($\left\langle CB\right\rangle$) and dichroism ($\left\langle CD\right\rangle$). Furthermore, the individual topological structure ($m,n$) of optical anisotropy maps ($G$) can be discerned at each phase section ($\theta$) of the blood film object field. The coordinate heterogeneity of $G\left(m,n\right)$ distributions can be explained by the specificity of processes involving supramolecular spatial-angular crystallization of protein molecules and blood film dehydration. Lastly, the average level and range of optical anisotropy parameters ($G\equiv LB,\left\langle CB\right\rangle,LD,\left\langle CD\right\rangle$) exhibit an increase with the increment of physical $h_i$ and effective $h_i^*$ depths in the polycrystalline blood film. This is attributed to the fact that the increase in physical depth ($\theta\le0.7~rad$) corresponds to an enhancement in the degree of spatial-angular ordering of supramolecular protein networks ($LB\uparrow,LD\uparrow$) and the number of formed elements ($\left\langle C B\right\rangle\uparrow,\left\langle C D\right\rangle\uparrow$) in the polycrystalline blood film. Within the range of multiple scattering ($0.7~rad\le\theta\le1.4~rad$), this process is intensified for $h_i^*$.

\begin{figure*}[ht!]
\begin{center}
\begin{tabular}{c}
\includegraphics[width=1.6\columnwidth]{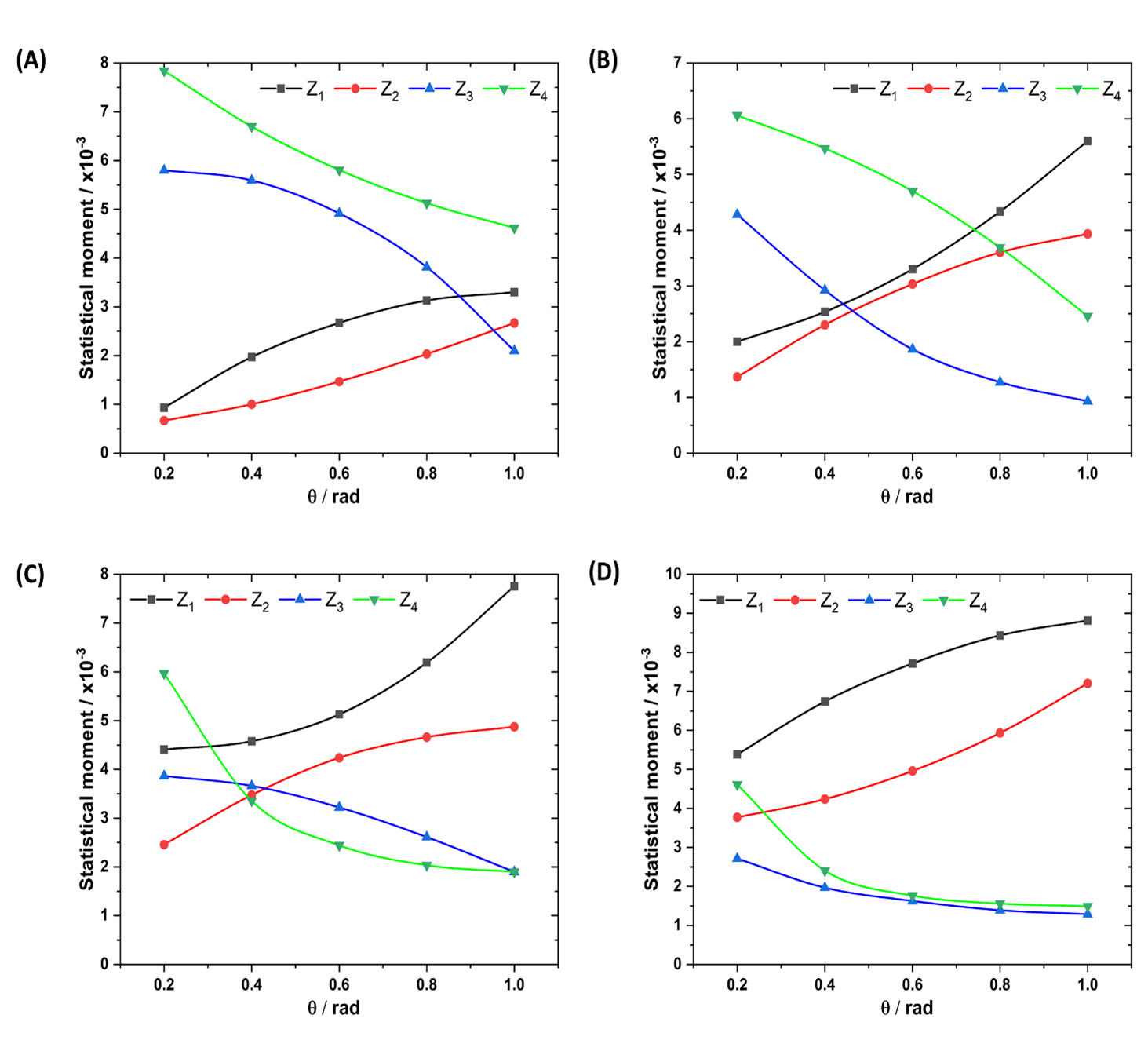}
\end{tabular}
\end{center}
\caption 
{\label{fig:6}
Phase-dependent magnitudes of the 1st (black, squares – $\times 10^{-3}$), 2nd (red, circles – $\times 10^{-3}$), 3rd (blue, upwards triangles), and 4th (green, downwards triangles) statistical moments characterizing the distributions of (A) linear birefringence, (B) circular birefringence, (C) linear dichroism, and (D) circular dichroism in a polycrystalline blood film from a healthy donor.} 
\end{figure*}

To quantitatively assess the transformation dynamics of algorithmically reconstructed optical anisotropy maps ($G\equiv LB,\left\langle C B\right\rangle,LD,\left\langle C D\right\rangle$) at each phase plane $\theta_k$, a statistical analysis is conducted according to (\ref{eq:29}). Figure \ref{fig:6} illustrates a series of ``phase" dependencies for the magnitudes of the 1st to 4th orders statistical moments ($Z_{i=1;2;3;4}\left(\theta\right)$). The analysis of the obtained data revealed two contrasting scenarios for the behavior of $Z_{i=1;2;3;4}\left(\theta\right)$. The first scenario involves a monotonic "phase" increase in the 1st and 2nd statistical moments, which characterize the mean and variance of the $G\left(\theta,m,n\right)$ distribution. The second scenario entails a decrement in the 3rd and 4th statistical moments, which characterize the skewness and kurtosis of optical anisotropy parameters. This behavior is attributed to the increased scattering multiplicity at $h_i^*(\theta\geq0.7~rad)$.

A multitude of optically anisotropic domains collectively contribute to the establishment of the average level of phase and amplitude anisotropy. Simultaneously, diverse geometric and concentration parameters within the fibrillar networks of proteins and optically active shaped elements of blood lead to an augmentation in the dispersion of linear and circular birefringence and dichroism in the polycrystalline blood film. The quantitative impacts of these processes are reflected in the values of the statistical moments. In the limit case, in accordance with the central limit theorem, the distribution $G\left(\theta\uparrow\right)\equiv LB,\left\langle C B\right\rangle,LD,\left\langle C D\right\rangle$ tends toward the normal distribution, and $Z_{3;4}\rightarrow0$.

\newpage

Comparing the changes in the 1st to 4th orders statistical moments, it was observed that skewness ($Z_3$) and kurtosis ($Z_4$) exhibit greater sensitivity to phase changes in the distributions of linear and circular birefringence and dichroism $G\left(\theta\right)$. This heightened sensitivity may be attributed to the fact that small variations in ($Z_2$) lead to larger changes in higher-order statistical moments. In the range $0.2~rad\le\theta\le0.7~rad$, the dynamic range of $\mathrm{\Delta}Z_{3;4}$ changes corresponding to linear birefringence and dichroism is 2.5-3 times, while for circular birefringence and dichroism, it is 3-4 times. Therefore, the layer-by-layer assessment of the polycrystalline structure in the phase shift range $0.2~rad\le\theta\le0.7~rad$ holds the potential for early-stage detection of oncological changes in the optical anisotropy of fibrillar networks of proteins and optically active shaped elements.

\subsection{Polycrystalline Blood Film Diagnosis}

\begin{table*}[!ht]
\centering
\caption{Optical parameters of polycrystalline blood film samples}
\label{tab:table-3}
\begin{tabular}{ccccc}
\hline
Statistical Parameter                     & \multicolumn{2}{c}{$\theta^*=0.65rad$} & \multicolumn{2}{c}{$\theta^*=0.45rad$} \\ \cline{2-5} 
 Difference & $LB$   & $\left\langle CB\right\rangle$       & $LD$         & $\left\langle CD\right\rangle $  \\ \hline
$\Delta Z_1$  & $0.06\pm0.002$     & $0.09\pm0.004$         & $0.05\pm0.002$        & $0.08\pm0.003$         \\ \hline
$\Delta Z_2  $ & $0.045\pm0.002$         & $0.075\pm0.003$         & $0.036\pm0.002$         & $0.064\pm0.003$         \\ \hline
$\Delta Z_3$ & $0.28\pm0.012$        & $0.35\pm0.014$         & $0.33\pm0.013$         & $0.49\pm0.018$         \\ \hline
$\Delta Z_4$ & $0.37\pm0.015$         & $0.52\pm0.023$         & $0.42\pm0.017$         & $0.063\pm0.027$         \\ \hline
\end{tabular}
\end{table*}

\begin{figure*}[ht!]
\begin{center}
\begin{tabular}{c}
\includegraphics[width=1.6\columnwidth]{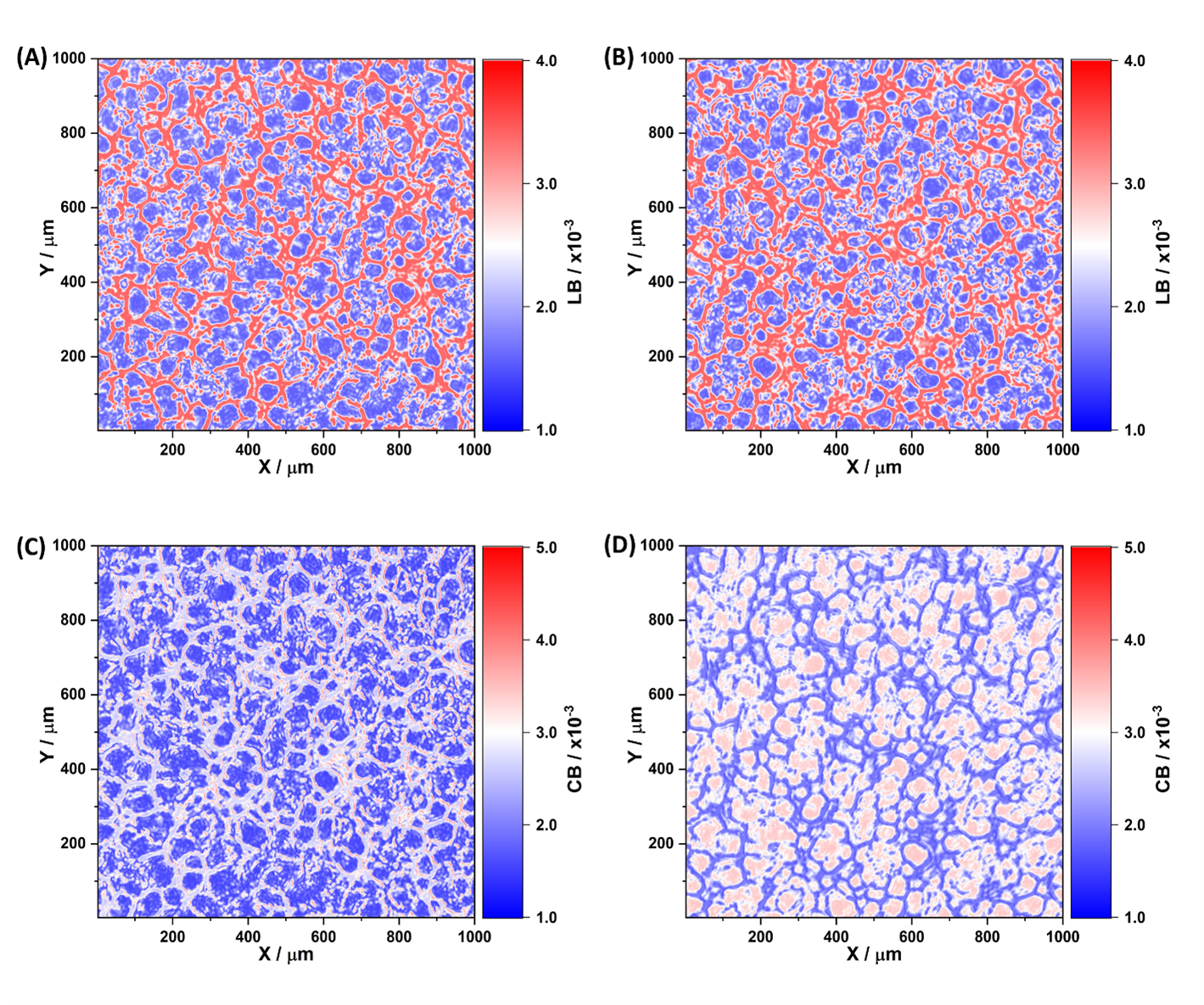}
\end{tabular}
\end{center}
\caption 
{\label{fig:7}
Maps of the (A-B) linear $LB(\theta^*=0.45rad,m,n)$ and (C-D) circular $CB(\theta^\ast=0.65rad,m,n)$ birefringence of polycrystalline blood films obtained for group 1 (A,C) and group 2(B,D). } 
\end{figure*}

The optimal phase planes for diagnostic purpose have been unidentified: $\theta^*( LB,\left\langle CB\right\rangle)=0.65~rad$ and $\theta^*(LB,\left\langle CB\right\rangle)=0.45~rad$. The optical anisotropy parameters obtained in these planes are illustrated in Figure~\ref{fig:6} and Figure~\ref{fig:7} for samples from group 1 (healthy) and group 2 (moderately differentiated prostate adenocarcinoma with a Gleason’s pattern scale of $3+3$).

In samples from group 2, a decrease in both the average level and fluctuations of linear birefringence and dichroism was observed. Conversely, an increase in both the average level and fluctuations of circular birefringence and dichroism was noted in the same group. From a physical perspective, these results can be linked to changes in the ratio between the concentrations of albumin and blood globulin proteins. It is well-documented~\cite{ushenko2012wavelet,ushenko2012mueller,ushenko2014two,ushenko2013spatial,ushenko2013spatiala} that early malignant processes are accompanied by an elevation in the concentration of optically active globulin molecules. The increased globulin concentration in group 2 contributes to the heightened magnitude of circular birefringence and dichroism compared to the healthy group.

The reduction in the concentration of albumin molecules, in turn, leads to a decrease in the level of linear birefringence and dichroism of supramolecular protein networks. These biological changes are reflected in the intergroup differences $\Delta Z_i$ of the statistical moments $Z_i$ characterizing the optical anisotropy maps of polycrystalline blood films from groups 1 and 2 (Table~\ref{tab:table-3}).

The 4th-order statistical moment, representing the kurtosis of the distributions of phase ($LB, \langle CB\rangle$) and amplitude ($LD, \langle CD\rangle$) anisotropy parameters in polycrystalline blood films, demonstrates remarkable sensitivity to early signs of an oncological state.

Table~\ref{tab:table-4} presents the sensitivity, specificity ($Sp$), and balanced accuracy ($Ac$) values for the early diagnosis of prostate cancer using the 3D layer-by-layer MM mapping method. These values are calculated based on the intergroup difference in the fourth-order statistical moment for each of the four optical anisotropy parameters. The results reveal an excellent level of balanced accuracy, indicating high levels of selectivity and specificity in the diagnostic approach.

\begin{table}[!ht]
\centering
\caption{Operational characteristics of the diagnostic power of the 3D MM method }
\label{tab:table-4}
\begin{tabular}{ccccc}
\hline
Operational                     & \multicolumn{2}{c}{$\theta^*=0.65rad$} & \multicolumn{2}{c}{$\theta^*=0.45rad$} \\ \cline{2-5} 
 Characteristic & $LB$   & $\left\langle CB\right\rangle$       & $LD$         & $\left\langle CD\right\rangle $  \\ \hline
$Se, \%$  & 86.1     & 94.4         & 88.9        & 97.2         \\ \hline
$Sp, \%  $ & 83.3       & 91.7         & 86.1         & 94.4        \\ \hline
$Ac, \%$ & 84.7        & 93.1         & 87.5       & 95.8        \\ \hline

\end{tabular}
\end{table}

A comparative analysis of diagnostic efficacy was conducted with three existing polarimetric methods, as outlined in Table~\ref{tab:table-5}. The considered methods are:

i. Azimuth-invariant polarization mapping of the distributions of polarization azimuth $\alpha(m,n)$ in the object field of the biological layer~\cite{ushenko2012wavelet,ushenko2012mueller,ushenko2013complex,ushenko2004laser,angelsky2010statistical,ushenko2013spatial,ushenko2013spatiala};

ii. Azimuth-invariant polarization mapping of the distributions of polarization ellipticity $\beta(m,n)$ in the object field of the biological layer~\cite{tuchin2015tissue,tuchin2010handbook,ushenko2012mueller,swami2013conversion,tuchin2016tissue,lu1996interpretation,deboo2004degree};

iii. MM ($R_{ik}(m,n)$) mapping of biological layers~\cite{ushenko2012mueller,ushenko2014two,ushenko2013diagnostics};

iv. This work: 3D MM reconstruction ($3D-LB, \left\langle CB\right\rangle, LD, \left\langle CD\right\rangle$) of the parameters of phase and amplitude anisotropy in biological layers in 3D.

\begin{table}[!ht]
\centering
\caption{Balanced accuracy of different laser polarimetry methods for differentiating partially depolarising ($\Lambda=40\%-45\%$) polycrystalline blood films from healthy donors and patients with highly differentiated adenocarcinoma }
\label{tab:table-5}
\begin{tabular}{ccc}
\hline
 & $\alpha(m,n)$   & $\beta(m,n)$   \\ \hline
$Ac, \%$  & 55-65   & 60-65   \\ \hline
 & $R_{ik}(m,n)$       & $3D-LB,\left\langle CB\right\rangle,LD,\left\langle CD\right\rangle$ (This work)      \\ \hline
$Ac, \%$ & 70-75       & 93-95  \\ \hline

\end{tabular}
\end{table}

An assessment of the diagnostic effectiveness of 2D and 3D polarization mapping methods for prostate tumor layers with varying optical thickness revealed that, for partially depolarizing polycrystalline blood films ($\Lambda=40-45\%$), the balanced accuracy of coordinate polarization methods ($\alpha, \beta(m,n)$) and MM mapping mostly falls below a satisfactory level. However, the accuracy of early differential diagnosis achieved through the 3D MM reconstruction method described in this work represents a significant improvement.

\subsection{From Bench to Bedside: Envisioning the Clinical Role of MM Mapping}

The presented analysis includes sensitivity ($Se$), specificity ($Sp$), and balanced accuracy ($Ac$) for the comparison of group 2 (moderately differentiated prostate adenocarcinoma, $3+3$ on Gleason’s pattern scale) and group 3 (poorly differentiated prostate adenocarcinoma, $4+4$ on Gleason’s Pattern scale) (see Table~\ref{tab:table-6}).

The results in Table~\ref{tab:table-6} indicate a high level of efficiency (ranging from $90.3\%$ to $95.8\%$) in diagnosing prostate tumors through MM mapping of polycrystalline blood films from patients with prostate adenocarcinoma at varying degrees of differentiation.
\\
\begin{table}[!ht]
\centering
\caption{Operational Characteristics of the Diagnostic Power of the 3D MM Method for Prostate Adenocarcinoma Stage Differentiation.}
\label{tab:table-6}
\begin{tabular}{ccccc}
\hline
Operational                     & \multicolumn{2}{c}{$\theta^*=0.65rad$} & \multicolumn{2}{c}{$\theta^*=0.45rad$} \\ \cline{2-5} 
 Characteristic & $LB$   & $\left\langle CB\right\rangle$       & $LD$         & $\left\langle CD\right\rangle $  \\ \hline
$Se, \%$  & 94.4     & 97.2         & 91.7        & 94.4         \\ \hline
$Sp, \%  $ & 91.7       & 94.4         & 88.9         & 94.4        \\ \hline
$Ac, \%$ & 93.1        & 95.8         & 90.3       & 94.4        \\ \hline
\end{tabular}
\end{table}

\newpage

\section{Conclusions}
In conclusion, our study employed a 3D MM reconstruction approach for multiparameter polarimetry studies on the polycrystalline structure of dehydrated blood smears. The investigation revealed method’s sensitivity to subtle changes in optical anisotropy properties resulting from alterations in the quaternary and tertiary structures of blood proteins, leading to disturbances in crystallization structures at the macro level at the very early stage of a disease. More specifically, the developed 3D MM diagnostic approach demonstrated discernible early cancer-related alterations in optical anisotropy properties. This included an examination of spatial distributions of linear and circular birefringence and dichroism in partially depolarizing polycrystalline blood films sourced from healthy tissues and cancerous prostate tissues across various stages of adenocarcinoma. Observable and quantifiable changes in the 1st to 4th order statistical moments, characterizing the distributions of optical anisotropy parameters, were identified in different ``phase" sections of the blood smear volumes.

Emphasizing the advantages of the presented diagnostic approach over traditional methods, we highlighted its cost-effectiveness and simplicity, requiring only a basic polarization-based optical setup without the need for reagents. Additionally, the analysis of dehydrated blood samples is prompt, providing express results compared to the time-consuming nature of biochemical analysis. Notably, during measurements, all parameters of the polycrystalline structure can be assessed simultaneously. An excellent accuracy ($> 90\%$) for early cancer diagnosis and differentiation of its stages is achieved, demonstrating the technique’s significant potential for rapid and accurate definitive cancer diagnosis compared to existing screening approaches. This pioneering work marks
an initial step toward the development of an advanced, practical, and cost-effective toolkit for expedited, minimally invasive cancer diagnosis, integrated with conventional blood tests.

\begin{acknowledgments}
Authors acknowledge the support from the National Research Foundation of Ukraine, Project 2020.02/0061 and Project 2022.01/0034; Scholarship of the Verkhovna Rada of Ukraine for young scientists-doctors of science; ATTRACT II META-HiLight project funded by the European Union’s Horizon 2020 research and innovative programme under grant agreement No.101004462, the Academy of Finland (grant projects 358200, 351068), the Leverhulme Trust and The Royal Society (Ref. no.: APX111232 APEX Awards 2021).
\end{acknowledgments}

\section*{Disclosures}
The authors declare no conflicts of interest related to this work. 

\section*{Ethics approval}
This study was conducted in accordance with the principles of the Declaration of Helsinki, and in compliance with the International Conference on Harmonization-Good Clinical Practice and local regulatory requirements. Ethical approval was obtained from the Ethics Committee of the Bureau of Forensic Medicine of the Chernivtsi National University and the Bukovinian State Medical University (Chernivtsi, Ukraine), and written informed consent was obtained from all subjects prior to study initiation.

\bibliography{main}% Produces the bibliography via BibTeX.

\end{document}